\documentclass[a4paper,12pt, twoside]{article}
\usepackage{amsmath,float,bm}
\usepackage{hyperref}
\usepackage{authblk,graphicx}
\usepackage{tikz}
\usepackage{tkz-tab}
\usepackage[total={6.5in,8.75in},
top=1.2in, left=0.9in, includefoot]{geometry}

\title{\textbf{Chaos in 2-d Bohmian Trajectories}}

\author{A.C. Tzemos \footnote{Corresponding Author: atzemos@academyofathens.gr} and G. Contopoulos}
\affil{Research Center for Astronomy and 
Applied Mathematics of the Academy of 
Athens - Soranou Efessiou 4, GR-11527 Athens, Greece}
\begin{document}
\maketitle

\begin{abstract}
We make a short review of the most general mechanism for the generation of chaos in 2-d Bohmian trajectories, the so called `nodal point-X-point complex' (NPXPC) mechanism. The presentation is based on  numerical calculations  made with Maple and  is enriched with new results on the details of the generation of chaos, and the form of the potential around the NPXPC.
\end{abstract}



\section{Introduction}
Bohmian Quantum Mechanics (BQM) \cite{Bohm,BohmII,holland1995quantum} is an alternative  interpretation of Quantum  Mechanics. In BQM the quantum particles of a system whose state is described by a usual wavefunction $\Psi$, i.e. a solution of the time dependent Schr\"{o}dinger equation 
\begin{equation}
-\frac{\hbar^2}{2m}\nabla^2\Psi+V\Psi=ih\frac{\partial \Psi}{\partial t},
\end{equation}
follow  certain deterministic trajectories in space-time, dictated by the so called Bohmian equations of motion 
\begin{equation}\label{bohmeq}
m\frac{d\mathbf{r}}{dt}=\hbar Im\left(\frac{\nabla \Psi}{\Psi}\right).
\end{equation}
We note that the Bohmian equations are of first order in time, in  contrast to the second order classical equations of motion. BQM has attracted a lot of interest in the last decades both from a theoretical and an experimental standpoint (see the reviews \cite{benseny, pladevall2012applied} and the paper \cite{kocsis2011observing}).

The highly nonlinear character of the Bohmian equations of motion (\ref{bohmeq}) makes BQM ideal for the study of order and chaos in quantum phenomena with all the techniques provided by the theory of classical dynamics. Bohmian order and chaos have been  studied in the last 25 years \cite{frisk1997properties,bialynicki2000motion, falsaperla2003motion,wisniacki2005motion,wisniacki2007vortex}. Much  work  has been done by the research group of the RCAAM of the Academy of Athens \cite{contopoulos2020chaos}. In particular our main contributions in BQM are:
\begin{itemize}
\item The development of the most general mechanism responsible for the production of chaos in 2-d and 3-d Bohmian trajectories.
\item The systematic study of ordered trajectories  
and their coexistence with chaotic trajectories in both 2-d and 3-d Bohmian systems.
\item The study of the  integrability of Bohmian trajectories.
\item The discovery of the phenomenon of partial integrability in the trajectories of certain 3-d Bohmian systems.
\item The study of the interplay between entanglement and chaos and its implications on a very important problem in BQM, which is  the dynamical approximation of Born's rule $P(t)=|\Psi(t)|^2$ in the case of initial distributions of particles with $P_0\neq |\Psi_0|^2$. These studies have been made in systems of entangled optical qubits which are of great importance in quantum technology applications.
\end{itemize}

Maple has helped us significantly in our analytical and numerical computations upon which the above results are based, as well as in their accurate depiction. Thus here we are going to review some of our basic results in 2-d systems and extend them with new calculations and figures made solely with Maple.

\section{The model}

The quantum harmonic oscillator (QHO) is the most well studied system in Quantum Mechanics. The knowledge of the complete set of its solutions in analytical form is very beneficial and that is why the QHO has been used in many works on Bohmian chaos.

In our case we work with a 2-d QHO of the form $H=\frac{1}{2}(\dot{x}^2+\dot{y}^2+x^2+c^2y^2)$, whose state is described by the wavefunction  \cite{parmenter1995deterministic}
\begin{equation}\label{wavefunction}
\Psi=\exp\left(-\frac{1}{2}(x^2+cy^2+i(1+c)t)\right)\left(1+ax\exp\left(-it\right)+b\sqrt{c}xy\exp\left(-(1+c)t\right)\right),
\end{equation}
where $a, b, c$ are real constants. We consider the non resonant case with incommensurable frequencies.
As it is well known, in BQM chaos appears close to  the nodal points of the wavefunction (where $\Psi_R=\Psi_{Im}=0$). In our case (\ref{wavefunction}) we have only one nodal point moving around the configuration space and its position can be found analytically as a function of time. It is given  by the equations
\begin{eqnarray}
&x_{N}=-{\frac {\sin \Big ((1+c)t \Big) }{a\sin \left( ct \right) }},\label{xnod}\\&
y_{N}=-{\frac {a\sin \left( t \right) }{b\sqrt{c}\sin \Big( (1+c)t\Big) }}.\label{ynod}
\end{eqnarray} 
The Bohmian equations corresponding to our model read:
\begin{eqnarray}
&\frac{dx}{dt}=-{\frac {\sin \Big ((1+c)t \Big) \sqrt {c}by+\sin \left( t \right) a
}{G}}\\&
\frac{dy}{dt}=-{\frac {\sqrt {c}bx \left( \sin \left( ct \right) ax+\sin \Big ((1+c)t \Big)  \right) }{G}},
\end{eqnarray}
where 
\begin{equation}
G=2\,\sqrt {c}ab{x}^{2}y\cos \left( ct \right) +c{b}
^{2}{x}^{2}{y}^{2}+2\,\sqrt {c}bxy\cos \Big ((1+c)t \Big) +{a}^{2}{x}
^{2}+2\,ax\cos \left( t \right) +1
\end{equation}

\begin{figure}[!ht]
\centering
\includegraphics[scale=0.22]{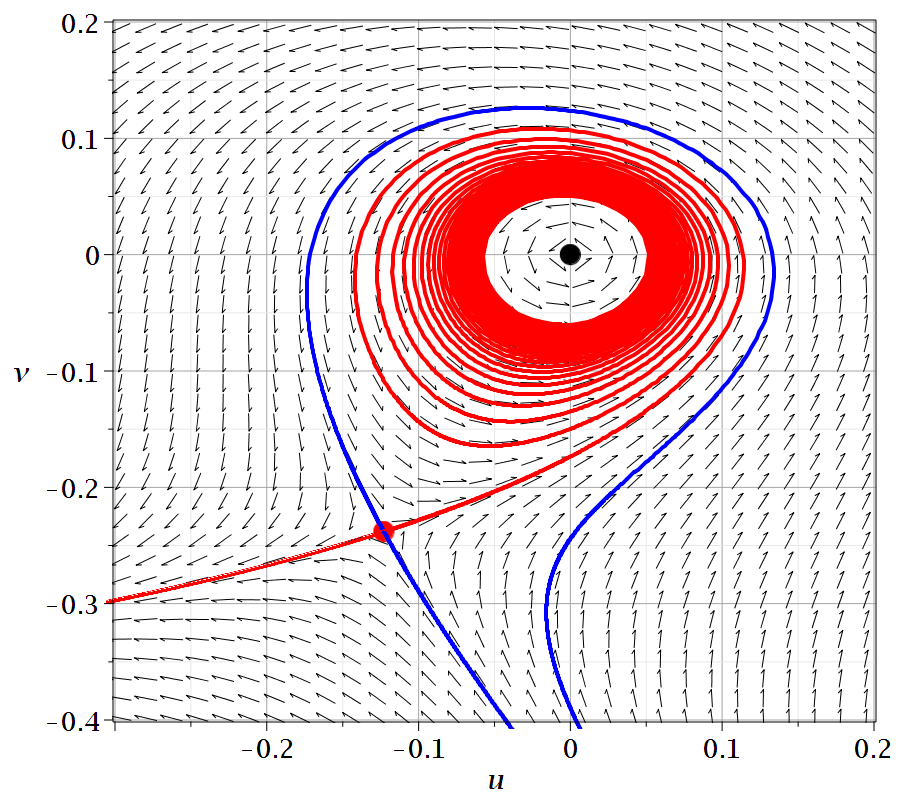}
\caption{The Bohmian flow in the neighbourhood of an NPXPC in the case $a=b=1$, $c=\sqrt{2}/2$ at time $t=1.01$, on the $(u,v)$ plane (red color: unstable asymptotic curves, blue color: stable asymptotic curves). The nodal point $N$ is an attractor ($A$) in this case.  }\label{101}
\end{figure}

In the frame of reference of the  nodal point ($u=x-x_{N},v=y-y_{N}$)  there is an unstable hyperbolic stagnant point, the 'X-point', which is defined as the non-trivial solution of the equations
\begin{equation}
du/dt=0,\quad dv/dt=0.
\end{equation}
As it is well known from the theory of dynamical systems\cite{Contopoulos200210}, from the unstable hyperbolic point emanate two stable and two unstable asymptotic curves pointing to opposite directions. The nodal point along with the X-point and its asymptotic curves form a characteristic geometrical form of the Bohmian flow, the `nodal point-X-point complex' (NPXPC).  A typical example is given in Fig.~\ref{101}, where we show the NPXPC at $t=1.01$. In order to observe the NPXPC one needs to:
\begin{enumerate}
\item Freeze the time $t$ at a certain value and calculate the position and the velocity of the nodal point at that time.
\item Transform the Bohmian equations of motion in the $(u,v)$ frame of reference of the moving nodal point, so they are written in the form 
\begin{equation}\label{dudt}
du/dt=F_1(u,v,;t), \quad dv/dt=F_2(u,v,;t).
\end{equation}
\item Create the field plot of (\ref{dudt}) around the nodal point.
\item Solve the system $du/dt=dv/dt=0$  in order to find the position $(u_{X}, v_{X})$ of the X-point.

\item Calculate the Jacobian matrix of the system (\ref{dudt}) at the X-point in order to find its eigenvalues and their corresponding eigenvectors. 
\item Introduce a fictitious time $s$ so that the equations at fixed time $t$ become
\begin{equation}\label{duds}
du/ds=F_1(u(s),v(s);t), \quad dv/ds=F_2(u(s),v(s),;t)
\end{equation}
\item Solve the system (\ref{duds}) for 4 initial conditions very close to the X-point and along its eigendirections in order to calculate its invariant curves. The unstable invariant curves, which correspond to the positive eigenvalue of the X-point and point away from it, must be integrated in positive time $s$. On the other hand, the stable invariant curves which correspond to the negative eigenvalue of the X-point and point towards it, must be integrated in negative time $s$.

\end{enumerate}

We note that the NPXPC is a dynamical geometrical structure, namely it changes as time progresses. Thus, it provides us with information about the form that the trajectories would acquire if the system was autonomous and described by the `fixed time equations' (\ref{duds}). However, when the velocity of the nodal point is small and the Bohmian flow  changes slowly in time, so does the NPXPC structure. Thus the trajectories in real time $t$ on the $(u,v)$ plane are very close to the trajectories in time $s$. This is the `adiabatic approximation'.

In Fig.~\ref{101} we observe that one unstable asymptotic curve (red color) forms a spiral approaching the nodal point, while the opposite asymptotic curve goes to infinity (on the left). On the other hand, the two stable asymptotic curves (blue color) start at the lower infinity and the left one reaches directly the X-point from below, while the other one surrounds the spirals around the nodal point and reaches the X-point from above.

The trajectories of two points that approach the X-point from below on either side of the left stable asymptotic curve, are deviated to the left and to the right with exponentially increasing distance. 

On the other hand, the trajectories that approach the X-point from above are also deviated to the left of the  figure or to the right. The latter belong to the set of trajectories that form spirals around the nodal point.

According to the NPXPC mechanism, every abrupt change of the direction of a trajectory during a scattering event with the X-point of a NPXPC is associated with a shift of the local Lyapunov characteristic number (the so called stretching number \cite{Contopoulos200210}). The cumulative action of many such scattering events results in the saturation of the maximal Lyapunov characteristic number at a positive value and thus to the emergence of chaos.

Finally, the trajectories coming from below between the two stable asymptotic curves approach asymptotically the nodal point. However, the proportion of such trajectories is small, because the separation between the two stable  asymptotic curves below the central part is also small. 

The NPXPC mechanism was first developed for 2-d Bohmian trajectories \cite{efthymiopoulos2007nodal,efth2009}  and then extended in the 3-d case \cite{tzemos2018origin}.

\section{Bohmian vortices: spiral motion around the nodal point}
In \cite{efthymiopoulos2007nodal,efth2009}
it was shown that  if  we introduce polar coordinates $(R, \phi)$, so that $u=Rcos(\phi), v=R\sin(\phi)$, we find that close to $N$ and in the lower approximation
\begin{equation}\label{spiral}
\frac{dR}{d\phi}=\langle f_3\rangle R^3,
\end{equation}
where 
\begin{align}
\langle f_3\rangle\nonumber=&\Bigg(\frac{1+b^2cx_{N}^4}{4b\sqrt{c}x_{N}^4\sin\Big((1+c)t\Big)}\Bigg)\times\\&
\Bigg(\frac{1-b^2cx_{N}^4}{1+b^2cx_{N}^4}x_{N}\dot{x}_{N}+\frac{\dot{x}_{N}\dot{y}_{N}(b^2cx_{N}^4-1)}{b\sqrt{c}\sin\Big((1+c)t\Big)}-x_{N}^2(\dot{x}_{N}^2-\dot{y}_{N}^2)\cot\Big((1+c)t\Big)\Bigg).\label{f3m}
\end{align}

Thus close to the nodal point the trajectories given by
\begin{equation}\label{spiral2}
R(\phi)=\frac{R_0}{\sqrt{1-2R_0^2\langle f_3\rangle(\phi-\phi_0)}}.
\end{equation}

\begin{table}[!ht]
\begin{tikzpicture}
\tkzTabInit[lgt=1,espcl=0.9,deltacl=0]
  { t/.8, $\langle f_3\rangle$ /.8, $\dot{\phi}$ /.8, $N$ /.8}
  {,$a\atop 1.28$,1.30,1.84,$b\atop 2.41$,2.47, $c\atop 2.75$,3.30,$d\atop 3.43$,3.68,$ 3.9428$,$e\atop 3.9431$,4.44,$f \atop  5.16$,5.52,} 
\tkzTabLine {,-,,+,,+,+\infty,+,,-,,-,,+,,+,,-,-\infty,-,,-,,+,z,-,,+,+\infty,} 
\tkzTabLine {,+,,+,,+,z,-,,-,,-,,-,,-,,-,z,+,,+,,+,,+,,+,z,}
\tkzTabLine {,A,,R,,R,,A,,R,,R,,A,,A,,R,,A,,A,,R,,A,,R,}
\end{tikzpicture}
\caption{Characteristic times and types of the nodal point. A particular transition appears at $t=4.44$ when $sin(ct)=0$. Then $\langle f_3\rangle$ changes sign but $\dot{\phi}$ does not change sign. Thus $N$ changes form from repeller to attractor.}\label{table1}
\end{table}

\begin{figure}[!ht]
\centering
\includegraphics[scale=0.25]{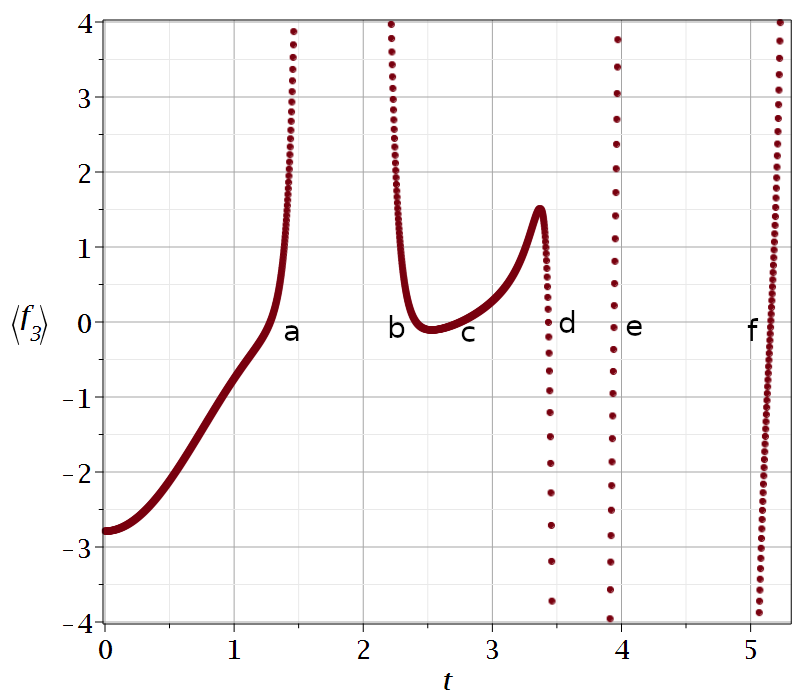}
\caption{The value of $\langle f_3\rangle$ as a function of time. It is zero  at the points $a, b, c, d, e, f$.}\label{f3}
\end{figure}

The trajectory \eqref{spiral2} is a spiral that approaches $N$ asymptotically. Namely if $\langle f_3\rangle<0$,  $R$ tends to zero when $\phi\to \infty$  and the nodal point is an attractor, and if $\langle f_3\rangle>0$, $R$  tends to zero  when $\phi\to-\infty$ (then $N$ is a repeller). Therefore, when $\langle f_3\rangle=0$ the nodal point changes its character from an attractor to  repeller or vice-versa (see Table~\ref{table1} and Fig.~\ref{f3}).

Furthermore, the theory specifies that $\dot{\phi}=\frac{d\phi}{dt}$ is proportional to $\sin\Big((1+c)t\Big)$ therefore $\dot{\phi}$ changes sign whenever $\sin\Big((1+c)t\Big)=0$, i.e. $t=\frac{k\pi}{1+c}, k=1, 2,\dots$ At these times we see from Eqs.~(\ref{ynod},\ref{f3m}) that $y_{N}=\pm\infty$ and $\langle f_3\rangle$ is also infinite. However $\langle f_3\rangle$ becomes infinite  also when $x_N$ is infinite (i.e. when $\sin(ct)=0$, therefore $t=\frac{k\pi}{c}, k=1, 2,\dots$). Finally, when the value of $\langle f_3\rangle$ changes sign as $t$ goes beyond a time $t=k\pi/c$,  $\dot{\phi}$ does not change its sign.

\begin{figure}[!ht]
\centering
\includegraphics[scale=0.216]{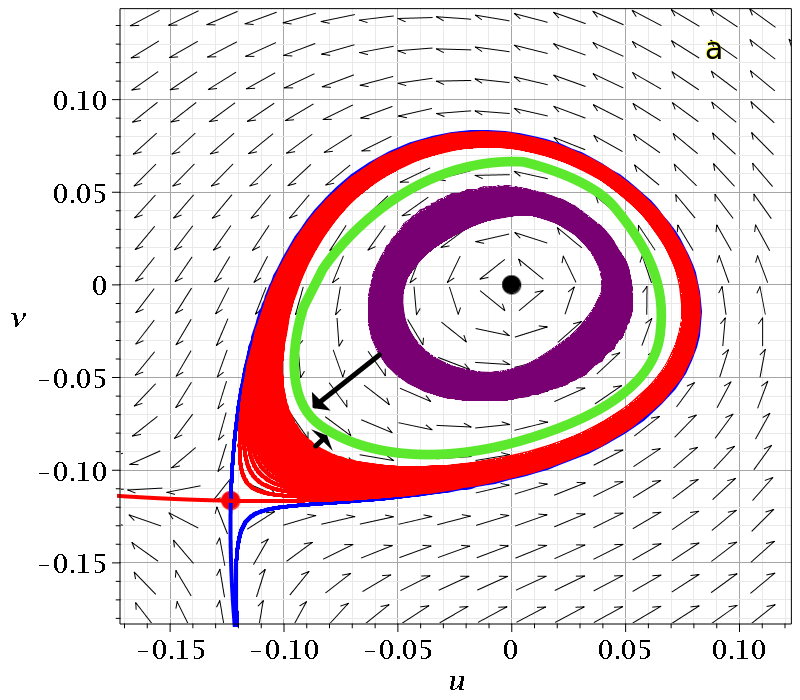}
\includegraphics[scale=0.22]{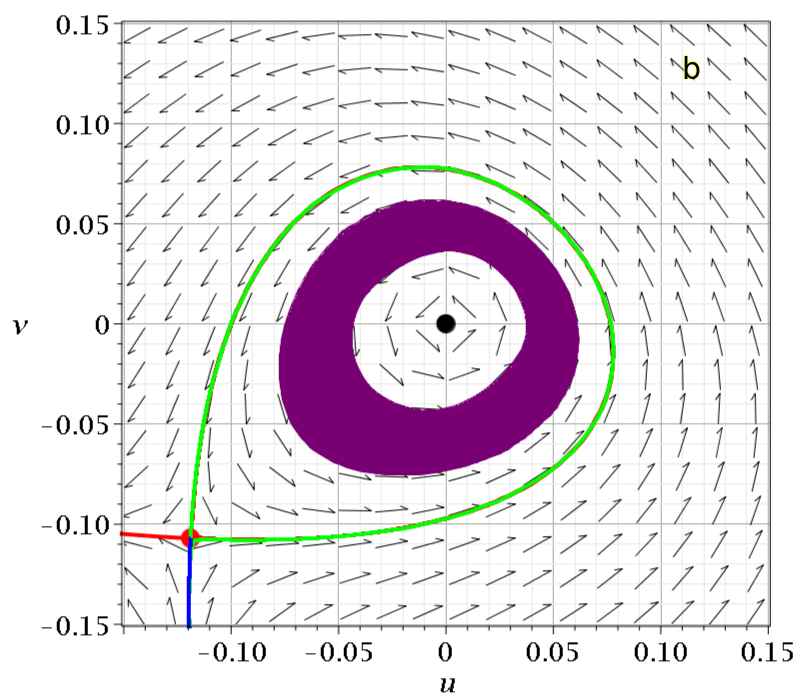}\\
\includegraphics[scale=0.22]{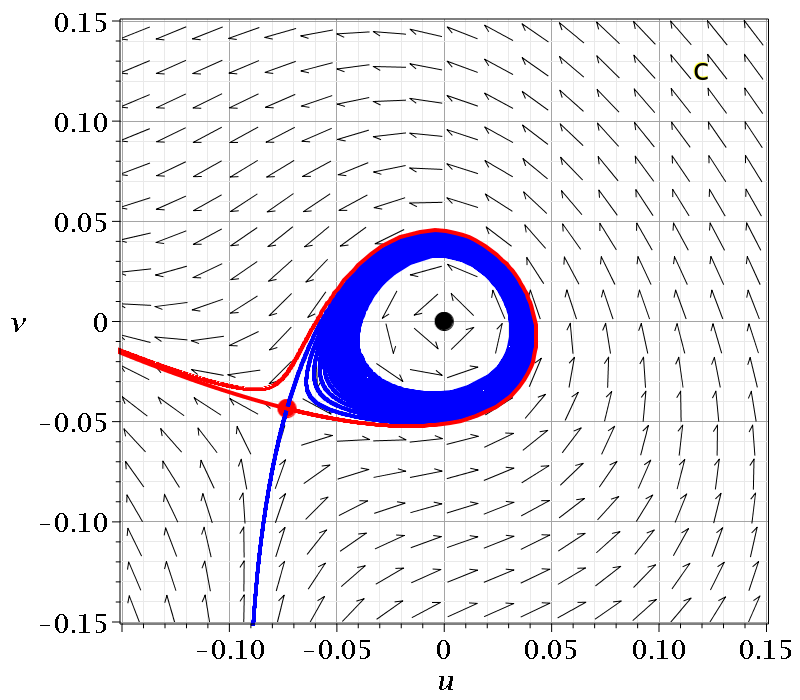}
\includegraphics[scale=0.22]{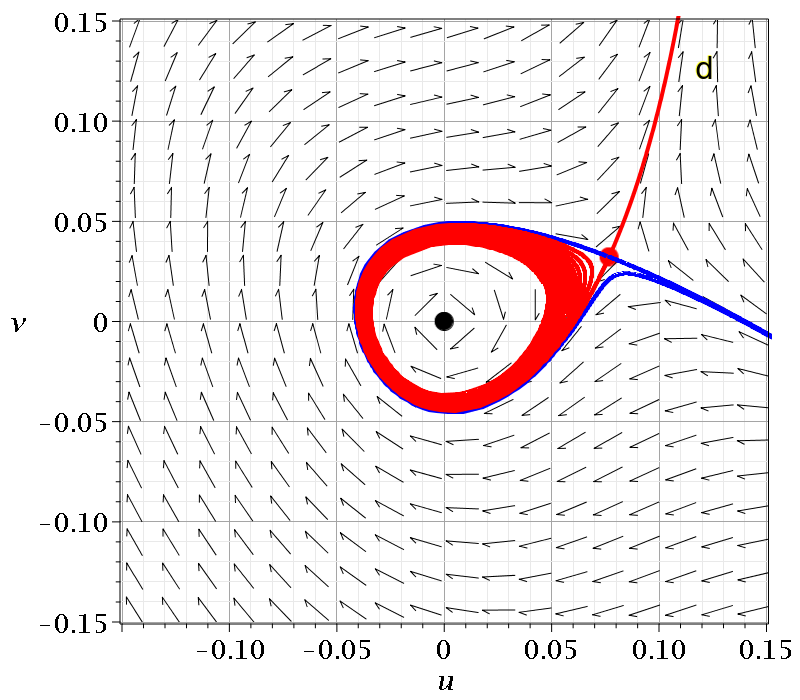}
\caption{The asymptotic curves: (a) At $t=1.29$  we observe a limit cycle  (green curve). The trajectories approach it (arrows)  both from outside and inside ($N$ is a repeller).  (b) At $t=1.303$ the limit cycle has reached the point $X$ and its asymptotic curves to the right and above.  (c) At $t=1.4$ the asymptotic curve  that approaches $N$ in negative time is stable (blue) ($N$ is a repeller). (d) At $t=2.3$ $N$ is an attractor and the asymptotic curve that approaches it is unstable (red).}\label{tetradaf}
\end{figure}

In Fig.~\ref{101} the nodal point $N$ is an attractor ($A$). But as time increases, $N$ becomes a repeller ($R$) at $t=1.28$. Beyond that time, a limit cycle is produced around $N$ and the trajectories close to $N$ go spiralling outwards counter-clockwise, approaching asymptotically the limit cycle, while the right unstable asymptotic curve and the nearby trajectories above it  on its left approach asymptotically the limit cycle spiralling inwards and counter-clockwise (Fig.~\ref{tetradaf}a). 

The limit cycle moves gradually outwards and reaches the X-point at $t=1.303$. Then the left unstable asymptotic curve joins the stable asymptotic curve  that surrounds $N$ and reaches the X-point from above (Fig.~\ref{tetradaf}b). Beyond that time the stable asymptotic curve above the point $X$ forms a spiral inwards in a negative time $s$ and reaches $N$ asymptotically. In positive time $s$ the nodal point continues to be a repeller ($R$). On the other hand, the unstable asymptotic curve to the right surrounds the spiralling trajectories around $N$ and escapes to the left close to the left unstable asymptotic curve (Fig.~\ref{tetradaf}c).

As time $t$ progresses the X-point approaches the nodal point and reaches it at $t=\frac{\pi}{1+c}\simeq 1.84$. At that time the nodal point goes to infinity ($y_{N}\to -\infty$) and jumps to $+\infty$. Beyond that time the nodal point comes closer again to the central region (near the origin  in the $(x,y)$ system) and $N$ is an attractor (Fig.~\ref{tetradaf}d). Since $\langle f_3\rangle$ is an even function of $\sin\Big((1+c)t\Big)$,  it is positive both before and after $t=1.84$, but as $\dot{\phi}$ is proportional to $\sin\Big((1+c)t\Big)$ the $\langle f_3 \rangle$ changes sign. The new form of the asymptotic curves beyond $t=\frac{\pi}{1+c}\simeq 1.84$ is shown in Fig.~\ref{tetradaf}d. Namely the asymptotic curve that spirals around $N$ approaching it asymptotically (spiralling inwards clockwise) is the lower unstable asymptotic curve, while the other unstable asymptotic curve goes upwards to infinity. The stable asymptotic curves come from infinity on the right. One reaches $N$ directly (in infinite time $s$) from the right, while the other reaches the X-point from the left after surrounding the spiralling trajectories around $N$ (Fig.~\ref{tetradaf}d). Beyond that time  we have similar sequences of changes (Table~\ref{table1}).

\section{Trajectories in the inertial frame of reference ($x-y$ plane)}

\begin{figure}[!ht]
\centering
\includegraphics[width=0.7\textwidth]{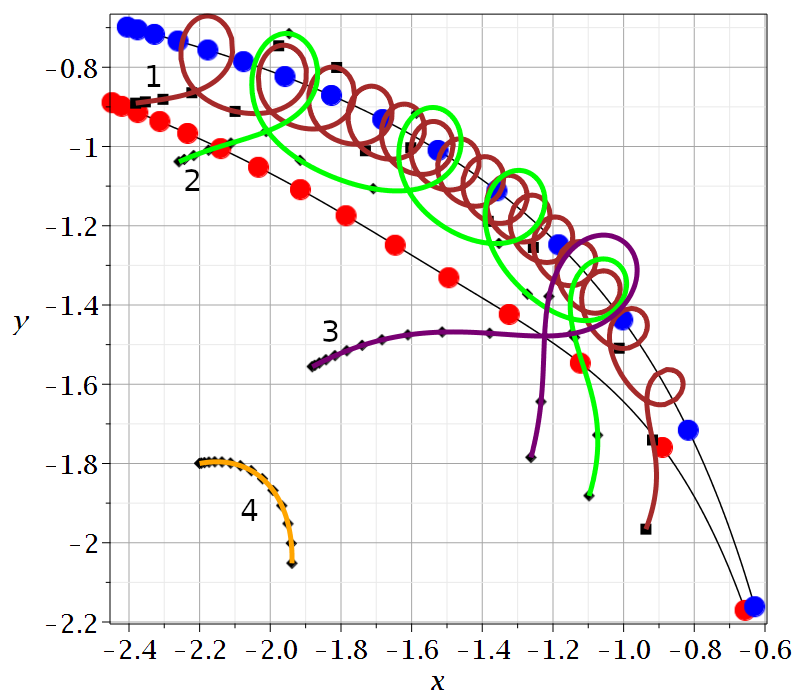}
\caption{Four Bohmian trajectories and the trajectories of the nodal point (blue points) and of the X-point (red points).
The initial conditions are: $x(0.1)=-2.381, y(0.1)=-0.891$ (crimson), $x(0.1)=-2.258, y(0.1)=-1.039$ (green) , $x(0.1)=-1.881, y=-1.556$ (purple) and $x(0.1)=-2.200, y(0.1)=-1.800$ (orange).
The  dots are at times $t=0.1, 0.2, 0.3\dots 1.5$.}\label{tksk}
\end{figure}

In Fig.~\ref{tksk} we draw four trajectories close to a moving nodal point and its associated X-point. The trajectory 1 (crimson) is closer to the nodal point and forms many spiral rotations around it. The nodal point and the X-point move on the $x-y$ plane for some time in an almost parallel way, but later on they approach each other. Finally these two points go to infinity ($y\to-\infty$) where they join.

The  trajectory 2 (green) is at a larger distance from the nodal point and forms a smaller number of spiral rotations around it.  Both trajectories  escape after some time from the neighbourhood of the moving NPXPC to large distances. The trajectory 3 forms a single loop around $N$, while the trajectory 4 (orange) is far away from $N$.

 In order to understand the escape mechanism we calculate the trajectories in the frame of reference of the moving nodal point ($u=x-x_{N},v=y-y_{N}$).

\begin{figure}[!ht]
\centering
\includegraphics[width=0.45\textwidth]{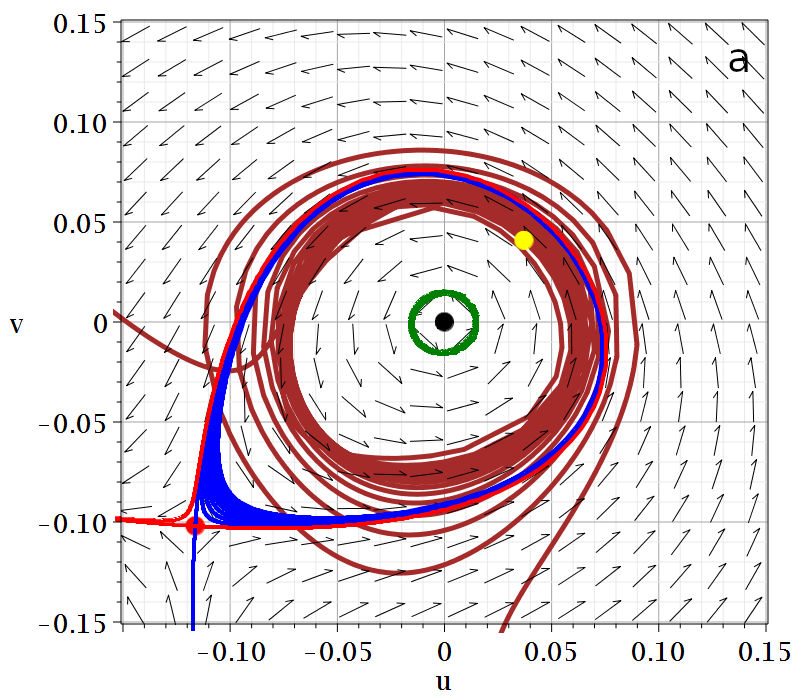}
\includegraphics[width=0.45\textwidth]{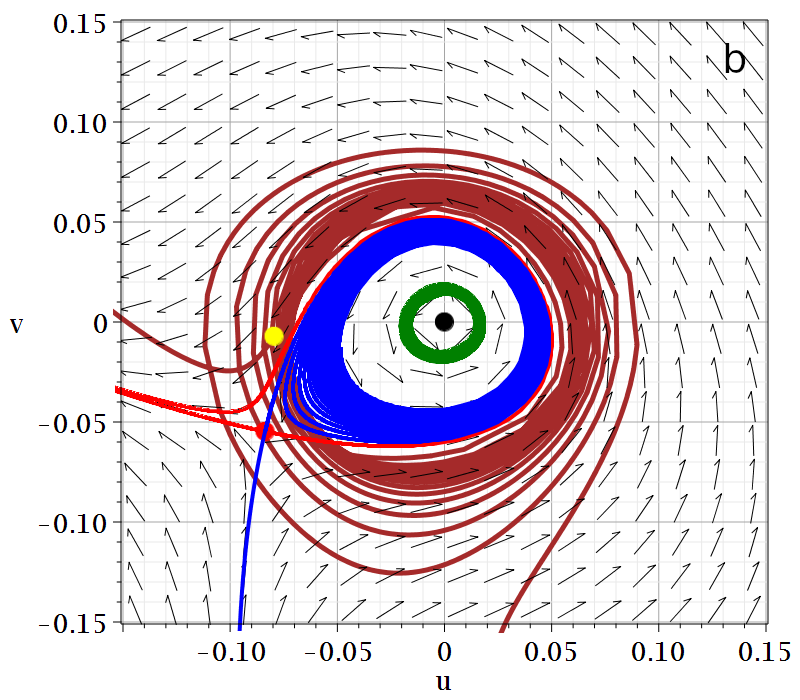}\\
\includegraphics[width=0.45\textwidth]{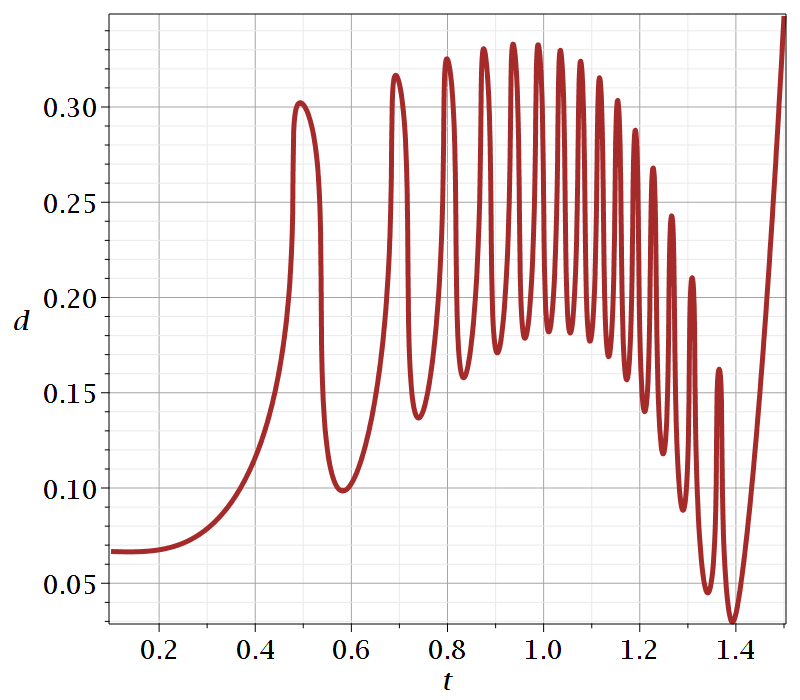}
\caption{The crimson trajectory on the plane $(u,v)$ together with the X-point and its asymptotic curves, stable (blue) and unstable (red). The yellow dot represents the position of the particle at the times  (a) $t=1.31$ and (b) $t=1.38$. (c) The distance between the Bohmian particle of the crimson trajectory and the X-point as a function of time.}\label{kitrino}
\end{figure}

In Figs.~\ref{kitrino}a,b we show the crimson trajectory  and the NPXPC on the $(u,v)$ plane at times $t=1.31$ (a) and $t=1.38$ (b) close to the escape time.  The domain of the spirals is surrounded by the unstable asymptotic curve from $X$ to the right and this domain is smaller in Fig.~\ref{kitrino}b, since the X-point is closer to the nodal point $N$. Furthermore, the trajectory is mostly inside the spiral domain in Fig.~\ref{kitrino}a but it is mostly outside the spiral domain in Fig.~\ref{kitrino}b. The exact position of the Bohmian  particle at that time is marked with a yellow dot. We see that this is well inside the spiral domain at $t=1.31$ (Fig.~\ref{kitrino}a) and outside of it at $t=1.38$ (Fig.~\ref{kitrino}b). A little later this point moves away from the domain of the spirals to the left, roughly parallel to the left asymptotic curve of the X-point.

\begin{figure}[!ht]
\centering
\includegraphics[width=0.45\textwidth]{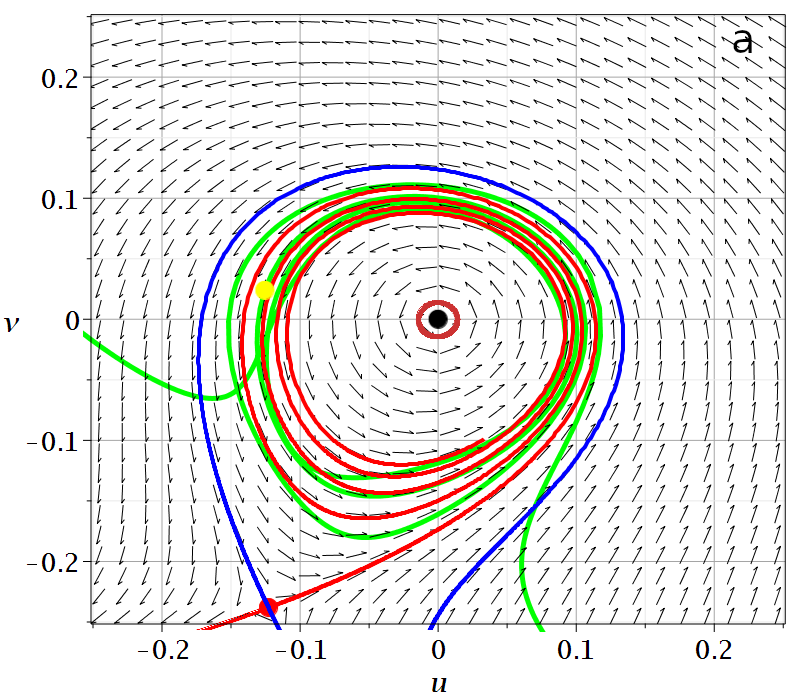}
\includegraphics[width=0.45\textwidth]{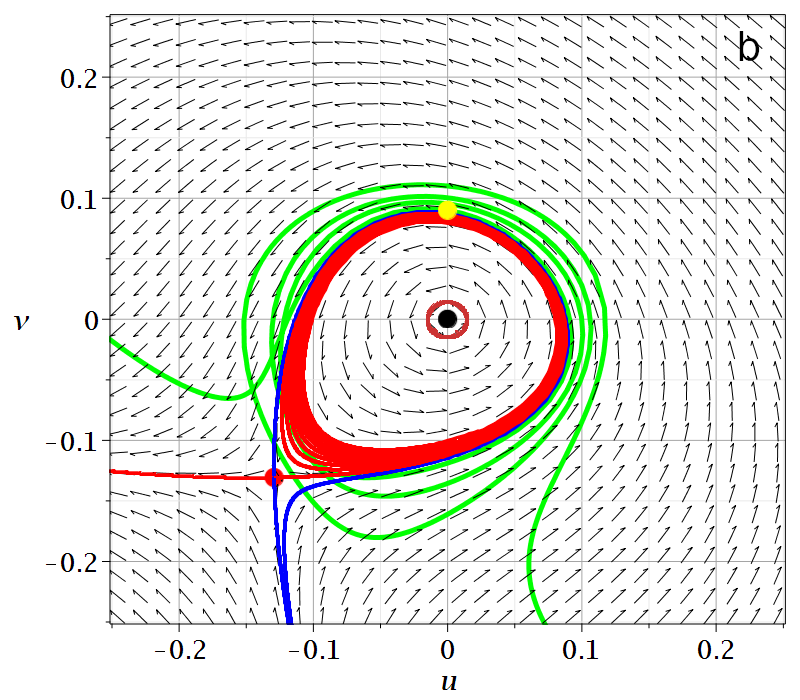}\\
\includegraphics[width=0.45\textwidth]{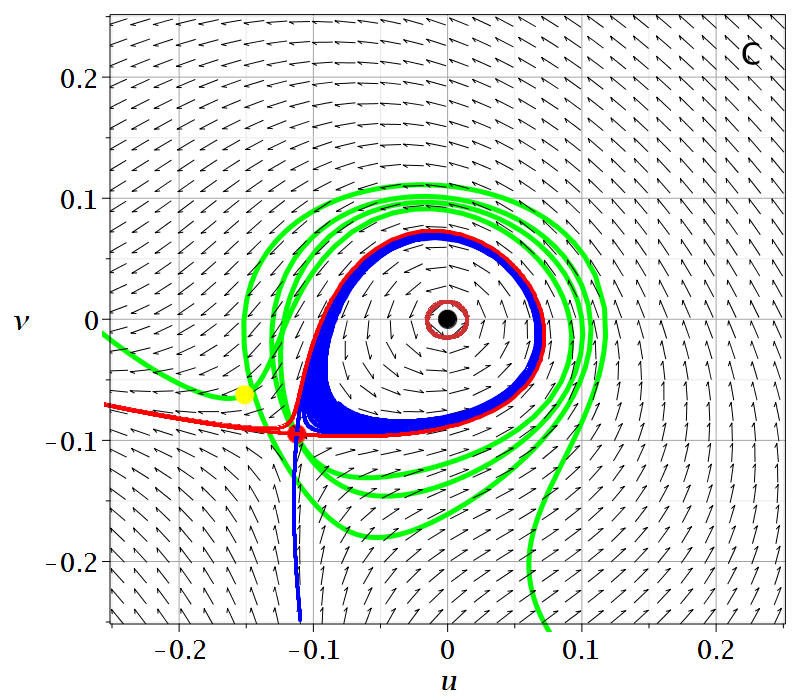}
\caption{As in Fig.~\ref{kitrino} for the green trajectory at the times (a) $t=1.01$, (b) $t=1.27$ and (c) $t=1.32$.}\label{kitrino2}
\end{figure}


\begin{figure}[!ht]
\centering
\includegraphics[width=0.5\textwidth]{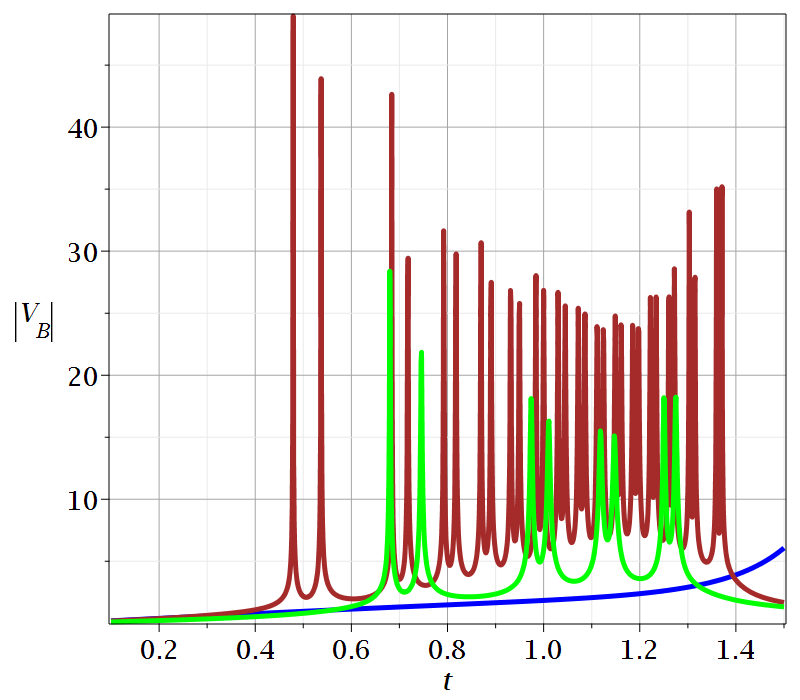}
\caption{The absolute values of the Bohmian velocities $|V_B|$ along the crimson and green trajectories.}\label{testvf}
\end{figure}

It is of interest to note that in Fig.~\ref{tksk} the moving point on the crimson trajectory escapes downwards  below the trajectory of the X-point   along an almost straight line, but in the frame $(u,v)$ (Figs.~\ref{kitrino}a,b) it escapes to the left turning by an angle about $\pi/2$.

The escape is due to the approach of the particle to the X-point. In Fig.\ref{kitrino}c we plot their distance as a function of time. We observe that after a number of oscillations (due to the spiral form of the trajectory), the distance comes to a minimum $d\simeq 0.03$ at  $t=1.39$. Beyond that time the distance grows indefinitely.

Similar results are found for the green trajectory (Figs.~\ref{kitrino2}a,b,c). In this case the spiral domain is defined by the blue stable asymptotic curve of the X-point. As time goes on the X-point approaches the nodal point $N$. In Fig.~\ref{kitrino2}a ($t=1.01$) the particle is well inside the spiral domain. In Fig.~\ref{kitrino2}b ($t=1.27$) it is just beyond this domain and in Fig.~\ref{kitrino2}c it has just escaped to the left along a trajectory almost parallel to the unstable asymptotic curve from $X$. 


Another characteristic of the trajectories close to the nodal point $N$ is that the velocities of the particles are very large  (Fig.~\ref{testvf} in comparison with Fig.~\ref{tksk}). When the particles are at their maximum distance above $N$ their velocity passes through a minimum. We have also a minimum between two approaches of the nodal point. Finally, after the escape of the particles from the NPXPC their velocities are small.

\begin{figure}[!ht]
\centering
\includegraphics[width=0.55\textwidth]{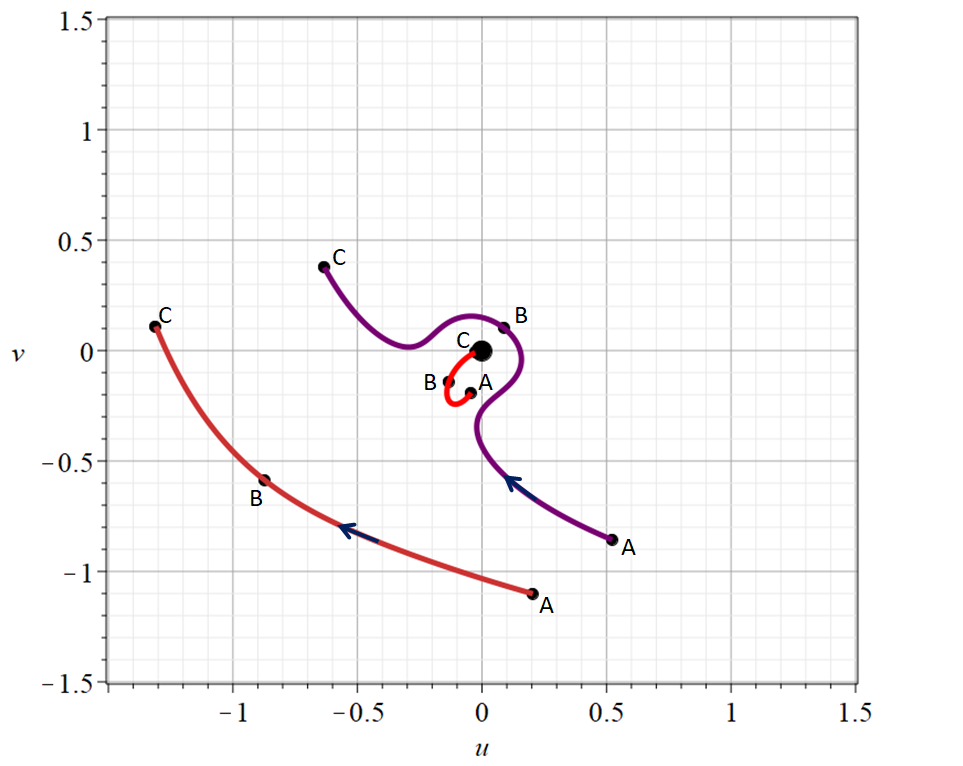}
\caption{The trajectories (3) (purple) and (4) (orange), and the trajectory of the X-point (red) on the $(u,v)$ plane where the nodal point $N$ is at the origin. The letters $A, B, C$ correspond at times: $t_A=0.1$, $t_B=1.256$  and $t_C=1.5$.}\label{type1_2}
\end{figure}

The third trajectory (purple) forms a single loop around $N$ and escapes immediately outwards, while the trajectories further away from the NPXPC (like the orange trajectory 4) are influenced by the NPXPC and are, in general, chaotic. In the frame of reference $(u,v)$ these trajectories are either surrounding the nodal point region (trajectory (3)) or beyond the X-point on the opposite side of $N$ (trajectory 4) (Fig.~\ref{type1_2}). In fact, the trajectories 3 and 4 are similar to those who undergo scattering processes of type I and type II, considered in \cite{efthymiopoulos2007nodal}. 

Regarding the trajectories of the points $N$ and $X$ in the inertial frame $(x,y)$, they are  given in Fig.~10. The nodal point and the X-point are initially at the points $A$ and $A'$. Then at a time $t=\frac{\pi}{1+c}\simeq 1.84$ both points go to $y=-\infty$ and then jump to $y=\infty$. Then they go to the points $B, B', CC'$ etc. They go to $y=-\infty$ at $t=3.68, 7.36...$ and  to $y=\infty$ at $t=5.52$ etc. Similarly, at $t=4.44$ the points $N$ and $X$ go to $x=-\infty$ and then jump to $+\infty$. At $t=9.52$ they go to $x=\infty$ and jump to $-\infty$ and so on. We note that at certain periods of time the point $X$ goes far away from $N$ (this happens around the times $t=2.8$ and $t=6.3$), but in general $X$ is close to $N$.

\begin{figure}[!ht]
\centering
\includegraphics[width=0.5\textwidth]{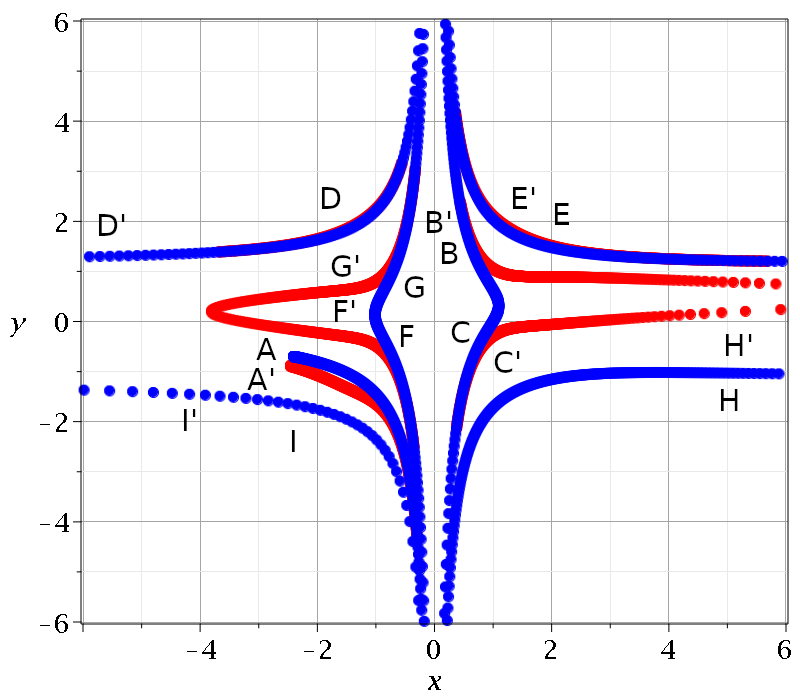}
\caption{The trajectories of the nodal point (blue) and the X-point (red) on the $(x,y)$ plane. The successive letters  $A-I$ (nodal point) and $A'-I'$ (X-point) correspond to times $t = 0, 2.25,  3.45,  4,  5.1, 5.85, 6.75, 8.0, 9.06$.}
\end{figure}

\begin{figure}[!ht]
\centering
\includegraphics[width=0.5\textwidth]{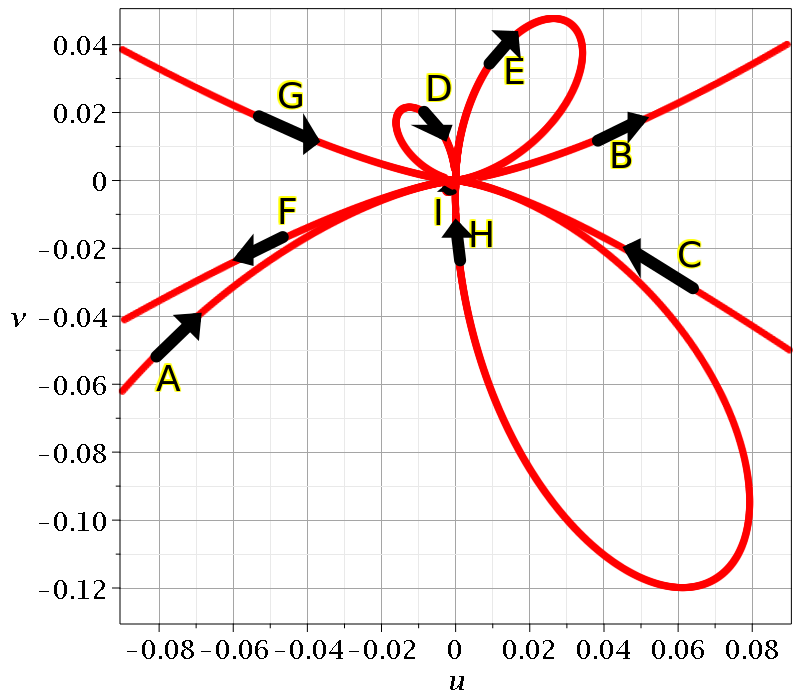}
\caption{Details of the trajectory of the X-point on the $(u,v)$  plane. The trajectory reaches the nodal point $(0,0)$ when $y_N=\pm\infty$ or $x_N=\pm \infty$. The letters $A-I$ correspond to times $t=0, 2.05, 2.65, 4, 5.1, 
6.3,  6.45, 8$.}\label{details}
\end{figure}

The relative position of $X$ with respect to $N$ is given in Fig.~\ref{details}. The position of the X-point $(u_x, v_x)$ when it is close to the nodal point $N$, is approximately given by \cite{efthymiopoulos2007nodal}

\begin{equation}
u_x=\frac{b\sqrt{c}}{L}\left(\sin\Big((1+c)t\Big)\right), \quad v_x=\frac{\dot{x}_{N}}{\dot{y}_{N}}\left(\frac{au_x^2\sin(ct)}{\sin\Big((1+c)t\Big)}-u_x\right),
\end{equation}

where $L$ is far from zero and $\frac{\dot{x}_{N}}{\dot{y}_{N}}$ is found by taking the derivatives of $x_{N}$ and $y_{N}$ from Eqs.~(\ref{xnod}) and (\ref{ynod}). Close to times $t=\frac{k\pi}{1+c}, k=1, 2,\dots$, $u_x$  is small and of order $\mathcal{O}(s)$, with $s=\sin\Big((1+c)t\Big)$, while $\frac{\dot{x}_{N}}{\dot{y}_{N}}$ is of order $\mathcal{O}(s^2)$ and $v_x$ is of order $\mathcal{O}(s^3)$. Therefore $v_x/u_x$ is of order $\mathcal{O}(s^2)$, i.e. the curve of the X-point approaches zero horizontally (Fig.~\ref{details}) while $x_{N}\to 0$ and $y_{N}=\pm\infty$. 

On the other hand, if $t$ is close to $t=\frac{k\pi}{c}$ we have (from Eqs.~(37) of \cite{efthymiopoulos2007nodal}) $\frac{\dot{x}_{N}}{\dot{y}_{N}}=\mathcal{O}(\tilde{s}^{-2})$ with $\tilde{s}=\sin(ct)$, $x_{N}=\mathcal{O}(\tilde{s}^{-1}), y_{N}=\mathcal{O}(1)$ (not close to zero or $\pm\infty$),  $L=\mathcal{O}(\tilde{s}^{-6})$,  therefore $u_x=\mathcal{O}(\tilde{s}^6), v_x=\mathcal{O}(\tilde{s}^4), v_x/u_x=\mathcal{O}(\tilde{s}^{-2})$.  As a consequence, the curve of $X$ in these cases reaches the point $N$ perpendicularly from above or below (Fig.~\ref{details}).


\section{The potential}
It is well known that the Bohmian trajectories can be written in a Hamiltonian form with a potential
\begin{equation}
V_{tot}=V_{cl}+Q,
\end{equation}
where
\begin{equation}
Q=-\frac{\hbar^2}{2m}\left(\frac{\nabla^2|\Psi|}{|\Psi|}\right)
\end{equation}
is the so called quantum potential. The total potential is quasi-periodic in time, even if $V_{cl}$ is time independent, because $Q$ contains terms with two frequencies, $1$ and $c$.

\begin{figure}[!ht]
\centering
\includegraphics[width=0.48\textwidth]{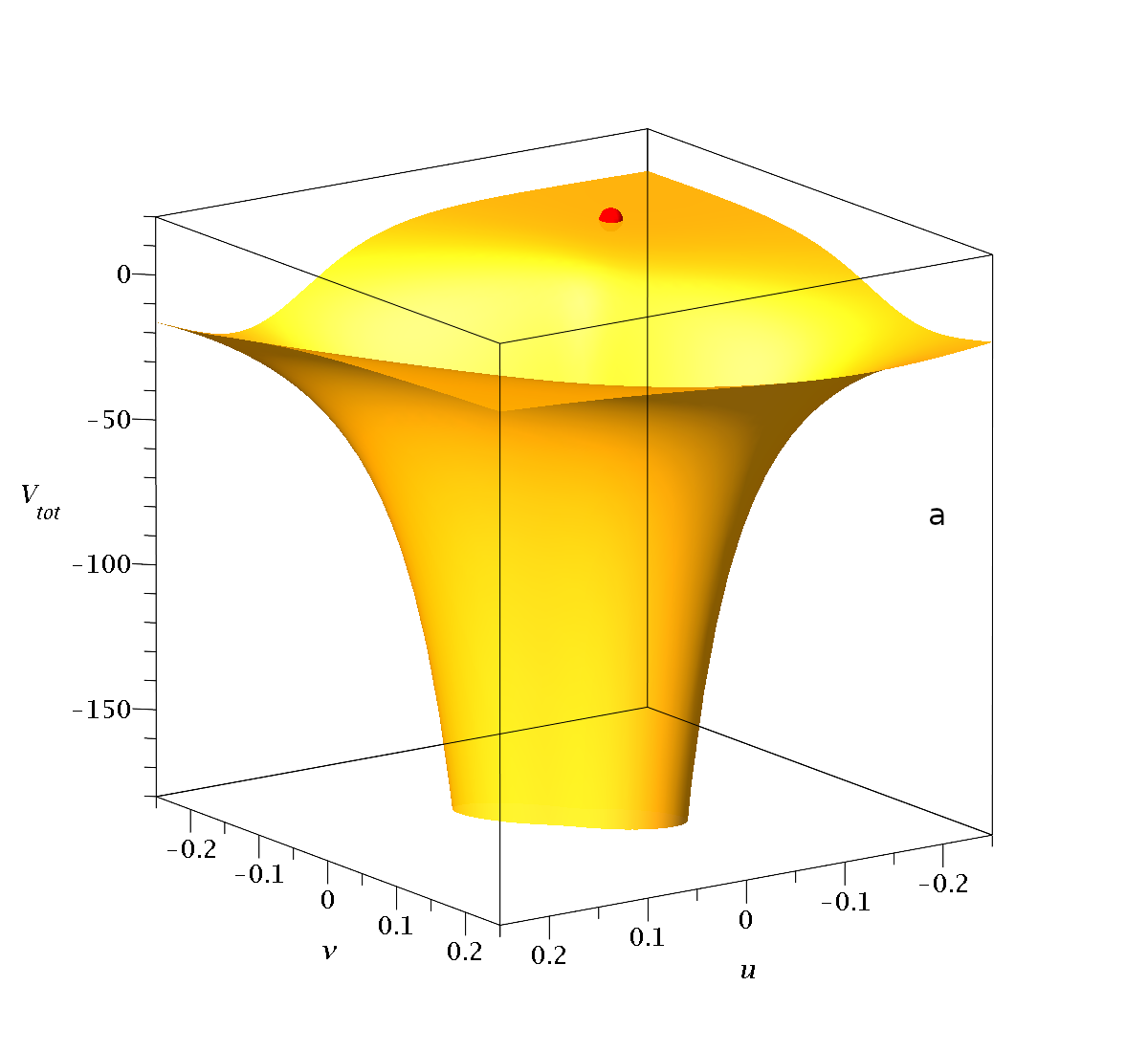}
\includegraphics[width=0.48\textwidth]{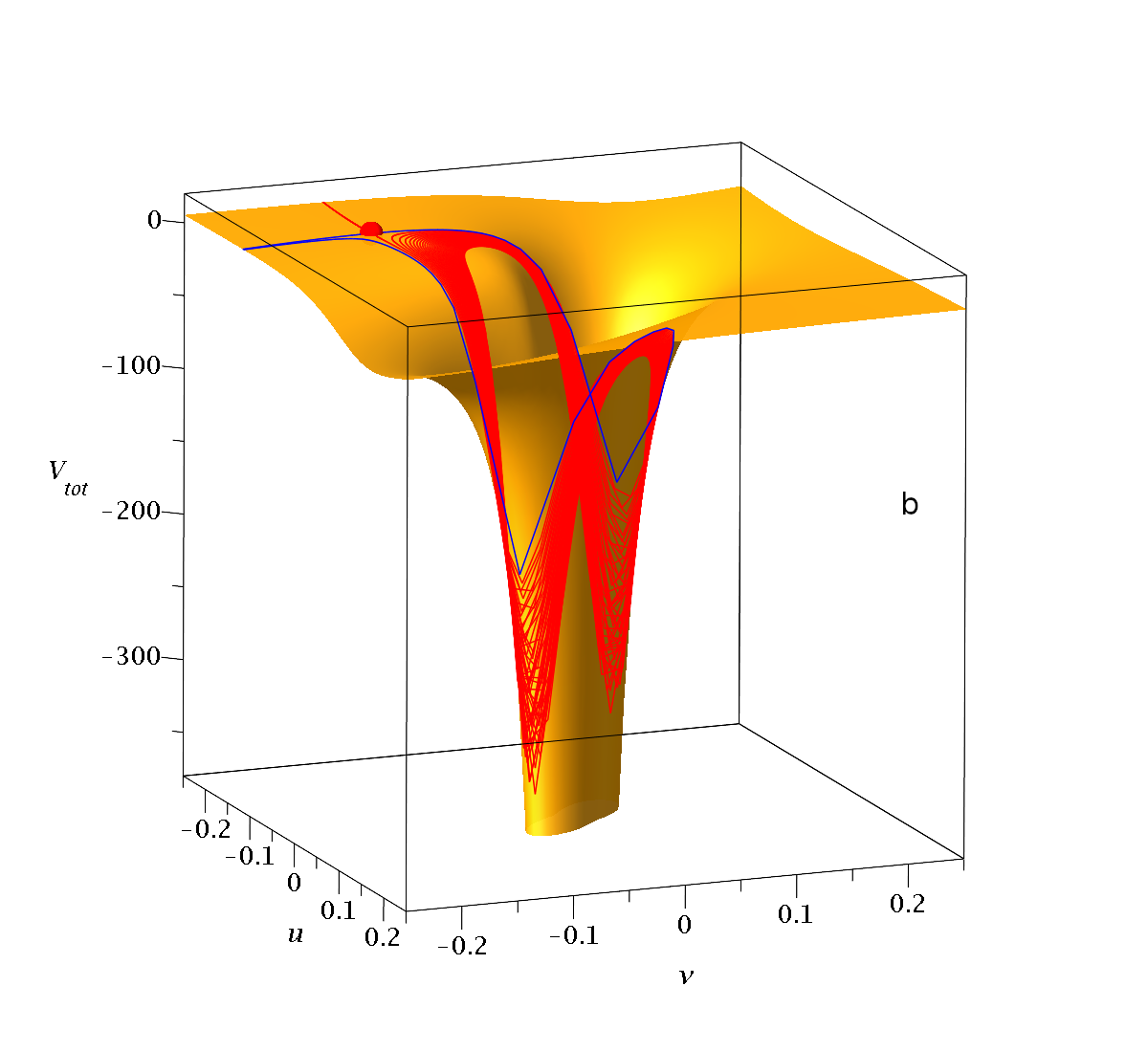}
\caption{a) The total potential and the X-point at $t=1.27$. b) Another view of the total potential  with the asymptotic curves of the X-point at the same time $t=1.27$. The red dot represents the X-point.}\label{Vol}
\end{figure}

\begin{figure}[!ht]
\centering
\includegraphics[width=0.45\textwidth]{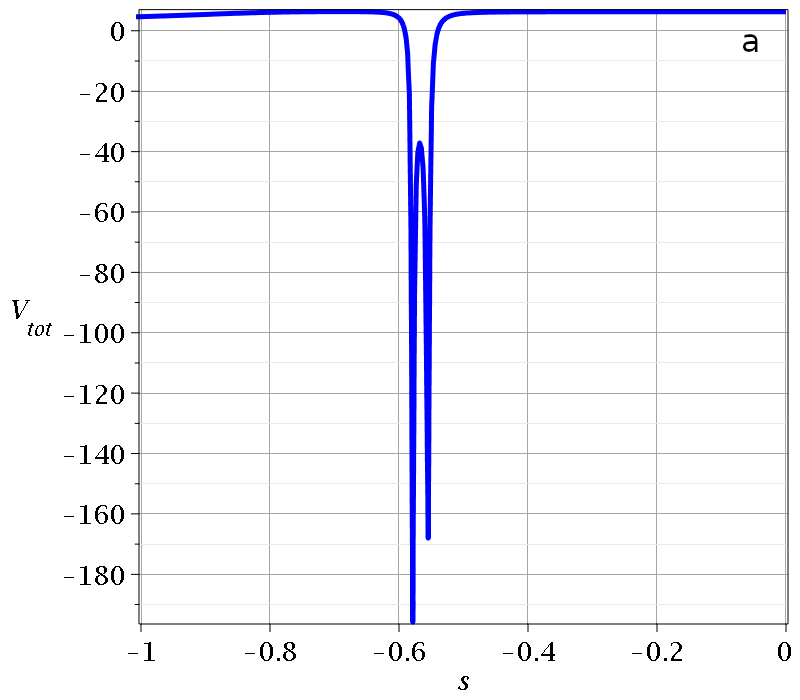}
\includegraphics[width=0.45\textwidth]{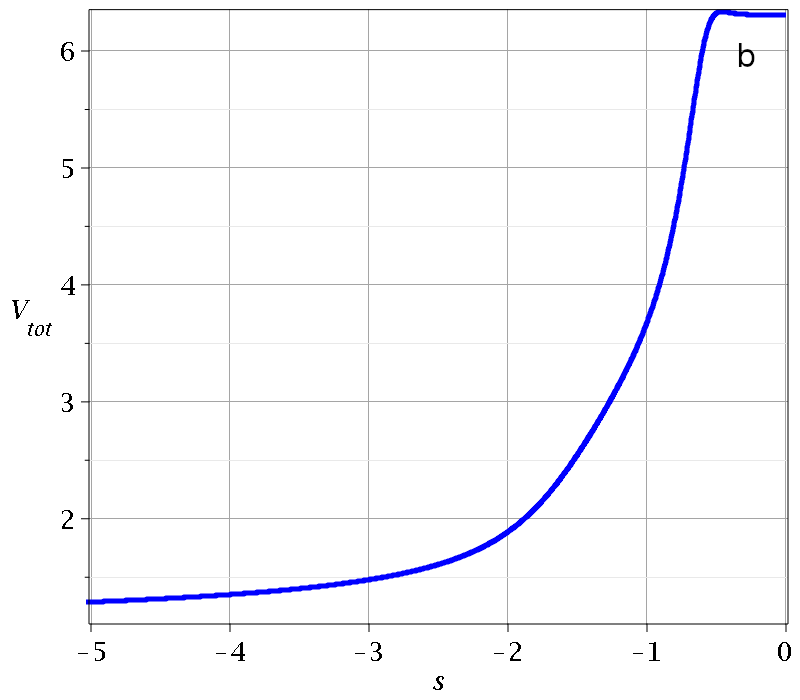}\\
\includegraphics[width=0.45\textwidth]{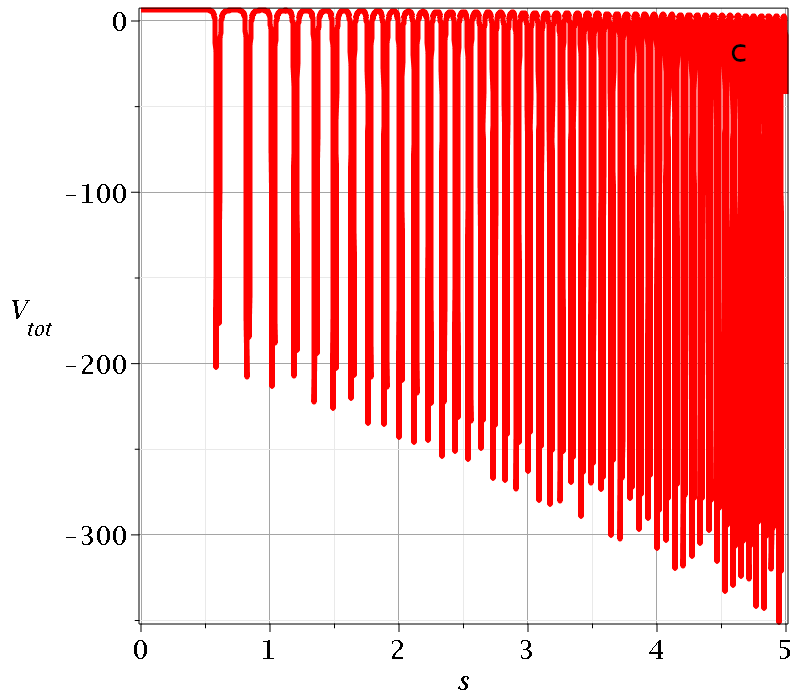}
\includegraphics[width=0.45\textwidth]{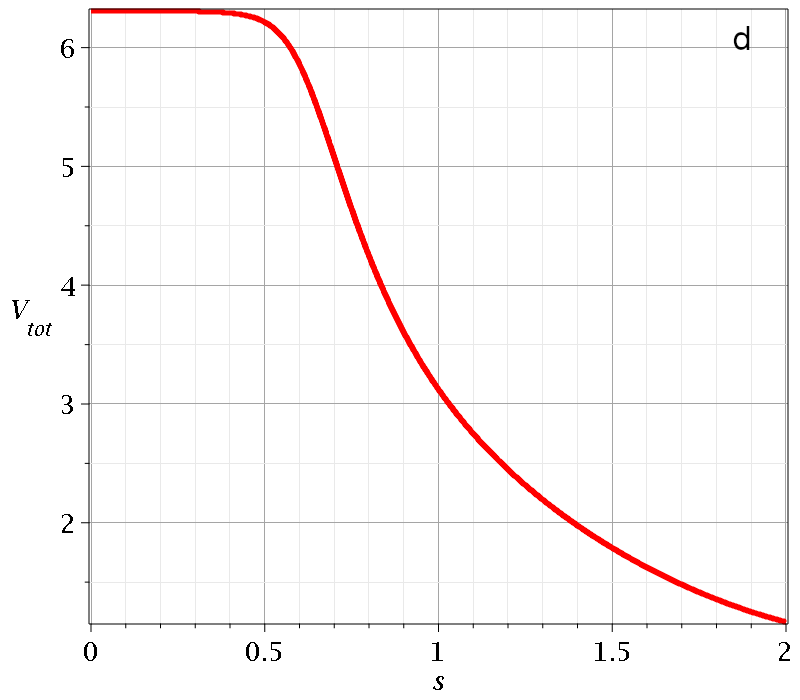}
\caption{The total potential in time $s$, along the stable asymptotic curves (upper panel) and the unstable  asymptotic curves (lower panel) of the X-point at fixed time $t=1.27$.}\label{asym}
\end{figure}
\begin{figure}[!ht]
\centering
\includegraphics[width=1.\textwidth]{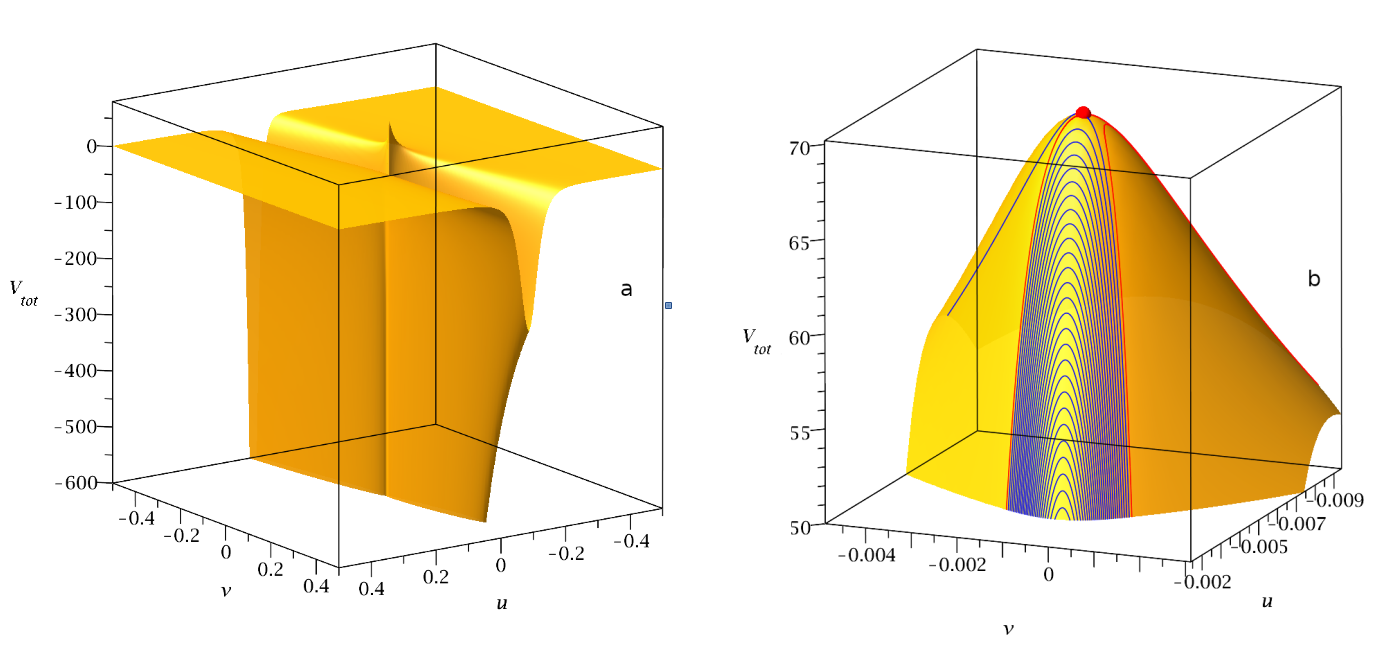}

\caption{a) The total potential $V_{tot}$ in the frame of reference $(u,v)$ of the nodal point at $t=1.6$. At that time the nodal point has  a large velocity which tends to infinity. b) The region close to the maximum of $V_{tot}$ where we have drawn also the stable and unstable asymptotic curves of the X-point (blue and red correspondingly).  }\label{3dzoom}
\end{figure}
In our case $V_{cl}=\frac{1}{2}(x^2+c^2y^2)$.
In Fig.~\ref{Vol}a we show the total potential in the $(u,v)$ frame at $t=1.27$.  We find that the X-point is close to the maximum of $V_{tot}$. In Fig.~\ref{Vol}b we have another view of the region and we have drawn also the asymptotic curves from the X-point. One stable (blue) curve from the X-point which is at a small distance from the nodal point  goes down twice and then it comes again close to $X$ and  escapes the region, close to the other blue curve from $X$. The  values of $V_{tot}$ along this and the other stable (blue) curve are shown in negative time $s$ in Figs.~\ref{asym}a,b. On the other hand, one red curve (unstable) makes an infinite number of downwards spirals (deeper and deeper), reaching the nodal point  at $s=\infty$. The corresponding values of $V_{tot}$ along this line and along the other red (unstable) line, in positive time $s$, are shown in Figs.~\ref{asym}c,d. We find that, as we approach the times $t=\frac{k\pi}{1+c}\, (k=1, 2,\dots)$ the maximum $V_{tot}$ increases. In fact in Fig.~\ref{3dzoom}a the values of $V_{tot}$ form almost a needle upwards, although the distance of the X-point from the nodal point is very small and the nodal point is at $V_{tot}=-\infty$. 

The X-point in the case $t=1.27$ is at ($u_x=-0.1292, v_x=-0.1308$) where $V_{tot}=6.308$. On the other hand, in the case $t=1.6$ the X-point is at $u_x=-0.0054, v_x=-0.0009$, i.e. it is much closer to the nodal point $N$. The value of $V_{tot}$ is $70.1703$, i.e. it is much larger than in the case $t=1.27$ and very close to the maximum $V_{tot}=70.1725$. But if we approach $N$ along the line $X-N$ at $(u=u_x/10, v=v_y/10)$ we find $V_{tot}\simeq -5.5\times 10^{3}$ and at $(u_x/100,v_y/100)$  it is $V_{tot}\simeq -6.6\times 10^{5}$. Thus the relation of $V_{tot}$ as we approach $N$ is very abrupt. In  Fig.~\ref{3dzoom}b we provide a detailed view   of the top of $V_{tot}$ close to the X-point together with its asymptotic curves. 

\begin{figure}[H]
\centering
\includegraphics[width=0.45\textwidth]{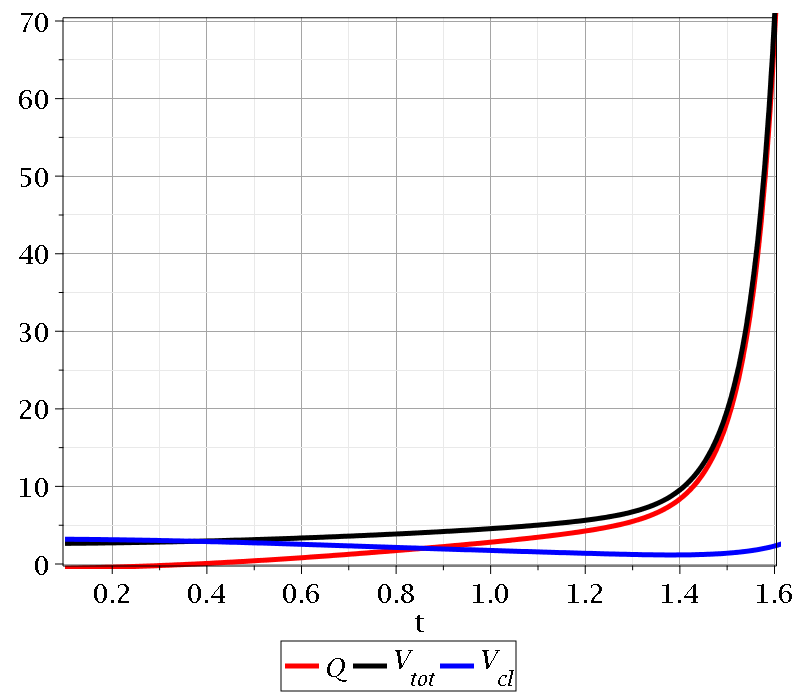}
\caption{The classical potential $V_{cl}$ (blue curve), the quantum potential $Q$ (red curve), and the total potential $V_{tot}=V_{cl}+Q$ (black curve) at the X-point as a function of time.}\label{dynamika}
\end{figure}

The values of $V_{tot}$ at the X-point increase continuously in time (Fig.~\ref{dynamika}) and they seem to go to $+\infty$ as $t\to\frac{\pi}{1+c} $. Thus the form of $V_{tot}$ tends to have two infinities, $+\infty$ and $-\infty$, at $t= \frac{\pi}{1+c}$. Before and beyond this time, the functions $Q$ and $V_{tot}=V_{cl}+Q$ have only one infinity $(-\infty)$ at $N$. Similar phenomena appear at higher critical values of $t$ (Table \ref{table1}).

\section{Calculations} 

Research in BQM is very demanding from a computational point of view. The character of Bohmian equations implies the coexistence of order and chaos, and chaos requires very accurate calculations. In particular, the emergence of chaos in the close neighbourhood of the nodal points, where the Bohmian velocities become very large, requires lengthy computations with many significant digits in order to get accurate results.  In fact, the numerical integration of Bohmian equations contains, in general, many short time intervals where the particle comes close to a moving NPXPC, and depending on the local geometry of the flow at that time, its direction changes abruptly or it gets trapped in a Bohmian vortex for a certain amount of time. The coexistence of different timescales in the time evolution of Bohmian trajectories makes their accurate calculation a very challenging and time  consuming process for most numerical integration schemes. 

We note that, in general, the wavefunctions have multiple nodal points wandering around the configuration space and scattering the trajectories. Here we used a rather simple wavefunction where we have only one nodal point whose position can be found analytically even by hand. This feature facilitates significantly all the calculations. In more complex cases the equations defining the nodal points are not exactly solvable and one needs to proceed with numerical techniques \footnote{ However, there are special cases with multiple nodal points, where their positions can still be found analytically \cite{tzemos2019bohmian,tzemos2020ergodicity,tzemos2021role}.}.
On the other hand, the calculation of the X-point is always numerical and becomes very tricky when the nodal points acquire large velocities, as they go far in the  the configuration space and tend to infinity. For example, the calculation of Fig.~\ref{details} required 25 significant digits.

The difficulty to calculate and depict accurate results increases significantly in the case of 3-d plots as those in Figs.~\ref{Vol} and  \ref{3dzoom}. In fact, in   Fig.~\ref{3dzoom} where the X-point is very close to the nodal point and the decrease of $V_{tot}$ between them is abrupt, we managed to get accurate results by working with 60 significant digits in Maple. 

\section{Conclusions}
BQM is a quantum theory where all the techniques of classical dynamical systems apply.  The Bohmian trajectories are either ordered or chaotic. According to the NPXPC mechanism the trajectories are chaotic if they approach an unstable hyperbolic point close to the nodal point (the X-point), which  has the same velocity with  the nodal point. The X-point has four asymptotic curves, two stable and two unstable. One of them forms a spiral towards the nodal point, while the  rest extend to infinity. Furthermore, the  nodal point is either an attractor or a repeller and sometimes close to the center of the NPXPC there is a limit cycle. 
The NPXPC evolves in time:
\begin{itemize}
\item As the  velocity of the nodal point increases abruptly (when it goes to infinity) the distance between $N$ and $X$ decreases and tends to zero. 
\item The character of the nodal point changes in time from attractor to repeller and vice-versa.
\item The trajectory which forms the spiral towards the nodal point changes in time, from stable to unstable and vice versa.
\end{itemize} 

In the present paper we presented the trajectories close to $N$ in the inertial frame of reference $(x,y)$  where they form loops around $N$ but later they escape to large distances from the NPXPC, after an approach to the X-point. We followed their evolution along with the trajectories of the points $N$ and $X$ which extend to infinity (where they join each other), at particular times. 

We then studied the  total potential, i.e, the sum of the classical potential $V_{cl}$ and the quantum potential $Q$ in the region of the NPXPC and found that the X-point is always very close to the positive maximum of $V_{tot}$ (while the nodal point is always at minus infinity). This is a new result and its generality needs to be further studied in the future.

All the calculations (symbolic and numerical) and their graphical presentation in this work have been made with Maple.

\section*{Acknowledgements}
This research was conducted in the framework of the program of the RCAAM of
the Academy of Athens “Study of the dynamical evolution of the entanglement and
coherence in quantum systems.”.

\bibliography{sample-base}

\end{document}